\documentclass[aps,prl,twocolumn,showpacs,superscriptaddress]{revtex4}
\usepackage{epsfig}
\usepackage{graphicx}
\usepackage{amsmath}
\begin{document}


\title{Fidelity Decay as an Efficient Indicator of Quantum Chaos}

\author{Joseph Emerson}
\email[Corresponding author: ]{jemerson@mit.edu}
\affiliation{Department of Nuclear Engineering}
\author{Yaakov S.~Weinstein} 
\affiliation{Department of Nuclear Engineering}
\author{Seth Lloyd} 
\affiliation{Department of Mechanical Engineering \\
Massachusetts Institute of Technology, 
Cambridge, Massachusetts, 02139} 
\author{D.G. Cory}
\affiliation{Department of Nuclear Engineering}

\date{\today}

\begin{abstract}
Recent work has connected the type of fidelity decay 
in perturbed quantum models to the presence of chaos 
in the associated classical models.  We demonstrate that a 
system's rate of fidelity decay under repeated perturbations 
may be measured efficiently on a 
quantum information processor, and analyze the conditions under 
which this indicator is a reliable probe of 
quantum chaos and related statistical properties of the unperturbed system.  
The type and rate of the decay are not dependent on the eigenvalue statistics of the unperturbed system, 
but depend on the system's eigenvector statistics in the  
eigenbasis of the perturbation operator. For random 
eigenvector statistics the decay is exponential with a rate fixed precisely by 
the variance of the perturbation's energy spectrum. 
Hence, even classically regular models can exhibit an 
exponential fidelity decay under generic quantum perturbations.
These results clarify which perturbations can distinguish 
classically regular and chaotic quantum systems.
\end{abstract}


\maketitle

Over the last two decades a great deal of insight has been achieved 
regarding the manifestations of chaos and complexity in 
quantum systems. We are interested in the problem of identifying 
such signatures in the context of quantum simulation on 
a quantum information processor (QIP). It is known that 
a QIP enables efficient simulation of the dynamics of a 
wide class of quantum systems \cite{Lloyd,ZalkaWeisner}. 
In the case of quantum chaos models, quantum simulation 
provides an exponential speedup over direct classical 
simulation \cite{Schack,Shep,Casati}.  
Recently, the quantum baker's map has been implemented 
using a nuclear magnetic resonance QIP \cite{Yaakov}. 
These developments highlight the importance of devising efficient QIP 
methods for the measurement of quantum chaos signatures 
and related properties of complex quantum systems.

Perhaps the most established signature of quantum chaos  
is given by the (nearly) universal correspondence 
between the eigenvalue \cite{BT77BGS84} and eigenvector 
\cite{IZR87KMH88HZ90} statistics 
of quantized classically chaotic systems and those of  
the canonical ensembles of random matrix theory (RMT). 
Unfortunately direct detection of these spectral signatures  
is algorithmically inefficient by any known technique. 
However, following the original observation of Peres \cite{Peres}, 
some recent work has demonstrated that, under 
sufficiently strong perturbation, 
the fidelity exhibits a characteristic exponential inthe case of 
classical chaotic systems
\cite{JP01Cucchietti01,Jacquod01,Casati01,Prosen02}. 
Below, we demonstrate that this charecteristic decay 
may be measured by an efficient algorithim, and analyze 
in detail how the observation of this decay may be 
applied as an indicator of canonical RMT 
statistics (quantum chaos)  in the {\em unperturbed} system. 

By first considering RMT models we show that 
Wigner-Dyson fluctuations in the system eigenvalue spectrum 
are not necessary to produce this characteristic 
decay. More importantly, 
we find that the canonical RMT statistics for the system 
eigenvector components, expressed  in the eigenbasis of the perturbation 
operator, are sufficient to produce the characteristic exponential decay. 
These observations are checked in the case of a dynamical model, 
where we demonstrate that an 
exponential fidelity decay can arise not only for a classically chaotic 
system, but also for classically regular system 
under a generic choice of perturbation. 
Our analysis implies specific restrictions that must be imposed on 
the choice of applied perturbation operator in order to extract 
useful information about the statistical signatures 
of quantum chaos in the unperturbed system. 

The aim of this Letter is to characterize certain static 
properties of a unitary map, $U$, by observing 
the rate of divergence between a fiducial Hilbert 
space vector evolved under 
this map, $|\psi_u(n) \rangle =  U^n |\psi(0) \rangle$, 
and the same initial vector evolved under this map but 
subject also to a sequence of small 
perturbations $| \psi_p(n) \rangle = (U_p U)^n |\psi(0) \rangle$, where 
$U_p = \exp(-i V \delta)$ is some unspecified perturbation operator 
and $n$ denotes the number of iterations. 
The fidelity,  
\begin{equation}
\label{eqn:fidelity}
O(n)  =  | \langle \psi_u(n) | \psi_p(n) \rangle |^2, 
\end{equation}
provides a natural indicator of this divergence. 
The value of $O(n)$ may be determined by an 
efficient algorithm on a QIP as follows. 
We start by preparing the fiducial state 
$| \psi(0) \rangle = U_o | 0 \rangle^{\otimes n_q}$, where $n_q$ is the number of 
qubits required to span the system's Hilbert space $N = 2^{n_q}$. 
As shown below, the choice of fiducial state is not critical and 
the computational basis states, which are simplest to implement, 
are a convenient set. 
After applying the sequence $(U^\dagger)^{n}(U_p U)^n$, 
the system register contains a final state $| \psi(n) \rangle$.  
The circuit implementation of $U$ requires only Poly($n_q$) operations 
for the simulation of a  wide class of quantum 
systems to arbitrary accuracy \cite{Lloyd,ZalkaWeisner,Shep,Schack,Casati}. 
Here we observe that $ | \langle \psi_p(n) | \psi_u(n) \rangle| = 
| \langle \psi(n) | \psi(0) \rangle|  =  
| \langle \psi_f | 0 \rangle |$,   
where the state $| \psi_f \rangle$ is obtained by time-reversing the 
initial state preparation $ |\psi_f \rangle = U_o^\dagger | \psi(n) \rangle$.  
The magnitude of $O(n)$ is then determined from sampling 
the population of the state $| 0\rangle $. 
The entire algorithm therefore scales as Poly($n_q$).
Recently, Jacquod and coworkers \cite{Jacquod01} observed that 
for $|\psi(0)\rangle 
= |v_o\rangle$, an initial eigenstate of $U$ 
with eigenphase $\phi_o$, the fidelity
relates to the local density of states 
(LDOS), $\eta(\phi_o-\phi_m') =  |\langle v_o | v_m' \rangle|^2$,  
via Fourier transform, 
\begin{equation}
\label{eqn:FT}
O(n) = | \sum_m \eta(\phi_o-\phi_m') \exp(-i (\phi_o -\phi_m') n) |^2. 
\end{equation}
In the above $|v_m'\rangle$ are eigenstates of the perturbed map 
$U_p U | v_m' \rangle = \exp(-i\phi_m') | v_m' \rangle $.
These observations locate the decay of $O(n)$
in an existing theoretical framework. 
In the non-perturbative regime $\sigma / \Delta \agt 1$, 
where $\sigma^2 = \delta^2 \overline{ V_{mn}^2}$ denotes  
a typical off-diagonal matrix element
and $\Delta$ is the average level spacing,  
previous work suggests that when the perturbed system is complex 
the LDOS is typically Lorentzian \cite{Wigner,FCIC96}, 
\begin{equation}
\label{eqn:LDOS}
\eta(\phi_o - \phi_m') \propto \frac{\Gamma}{ (\phi_o - \phi_m')^2 + (\Gamma/2)^2 }
\end{equation}
with width $\Gamma = 2 \pi \sigma^2 / \Delta$ determined by 
the Fermi golden rule (FGR). 
From (\ref{eqn:FT}) and (\ref{eqn:LDOS}) one expects 
the exponential decay, 
\begin{equation}
\label{eqn:fgr}
O(n) \simeq \exp(-\Gamma n). 
\end{equation} 
The onset of the exponential decay (\ref{eqn:fgr}) 
has been confirmed recently in a few  
classically chaotic models \cite{Jacquod01,Casati01}, 
though the rate is not always given by the golden 
rule \cite{Wisniacki02,JP01Cucchietti01}. 
However, in the case of integrable $U$ the situation is less clear  
since under some perturbations 
the LDOS is known to take on a Lorentzian shape \cite{FCIC96,BGI98}.

Below, we examine 
which statistical properties of the {\em unperturbed} system 
lead to the FGR decay, and how this characteristic decay  depends on the 
properties of the perturbation operator.  
This approach is motivated by the context of QIP simulation, in which 
the eigenbasis of this perturbation 
may be mapped onto an arbitrary basis of 
the simulated system $U$ \cite{Cory99}. 
Our choice of perturbation eigenvalue structure 
is motivated from the 
perspective of quantum control studies. 
Specifically, we consider  
\begin{equation}
\label{eqn:cbr}
U_p = \Pi_{j=1}^{n_q} \exp(- i \delta \sigma_z^j / 2 ),
\end{equation}
where $\sigma_z$ is the usual Pauli matrix
and (\ref{eqn:cbr}) therefore corresponds to a 
collective rotation of all the qubits by an angle $\delta$. 
Eq.~(\ref{eqn:cbr}) is a model of coherent far-field 
errors \cite{Fortunato}, and for this type of error model 
a better understanding of the fidelity decay is a 
subject of intrinsic interest. 

As a first test of the LDOS/FGR framework in the case of the 
qubit perturbation (\ref{eqn:cbr})  
we evaluate the fidelity decay for a map $U=U_{\text{CUE}}$ 
drawn from the circular unitary ensemble (CUE). 
These matrices form an established model for 
classically fully chaotic time-periodic systems 
(without additional symmetries) since these 
systems (almost always) exhibit the same characteristic 
(Wigner-Dyson) spectral fluctuations and  eigenvector statistics 
as the CUE \cite{Haake01}. 
The randomness of the system eigenvectors 
enables a system-independent estimate of the rate $\Gamma$ of the FGR decay.  
Since the components  of random eigenstates 
are distributed uniformly over 
the basis states and uncorrelated 
with the distribution of eigenvalues, 
the second moment of the matrix 
elements $V_{mn}$ may be directly evaluated, 
\begin{equation}
\label{eqn:Vmn}
\overline{V_{mn}^2} = \overline{\lambda^2}/ N  
\end{equation}
assuming $\overline{\lambda}=0$ and where 
$\overline{\lambda^2} = N^{-1} \sum_{i=1}^N \lambda_i^2$
denotes the variance of the eigenvalues of $V$.  
As a result, the rate of the FGR decay is determined by 
the eigenvalues of the perturbation, 
\begin{equation}
\label{eqn:Gamma}
	\Gamma = \delta^2 \overline{\lambda^2}, 
\end{equation}
where we have used $\Delta = 2 \pi /N$. 
For the qubit perturbation (\ref{eqn:cbr}) the variance of 
the eigenvalues has a simple form,
\begin{equation}
\label{eqn:bin}
	\overline{\lambda^2 } = \frac{1}{N}
\sum_{k=0}^{n_q} 
\left(\frac{2k - n_q}{2} \right)^2  C^{n_q}_{k}, 
\end{equation}
where 
the $C^{n_q}_k$ are binomial coefficients. 
Using our RMT estimate (\ref{eqn:Gamma}), for $n_q=10$, the rate is, 
\begin{equation}
\label{eqn:Gammacbr}
	\Gamma = 2.50 \; \delta^2.
\end{equation}



\begin{figure}
\epsfig{file=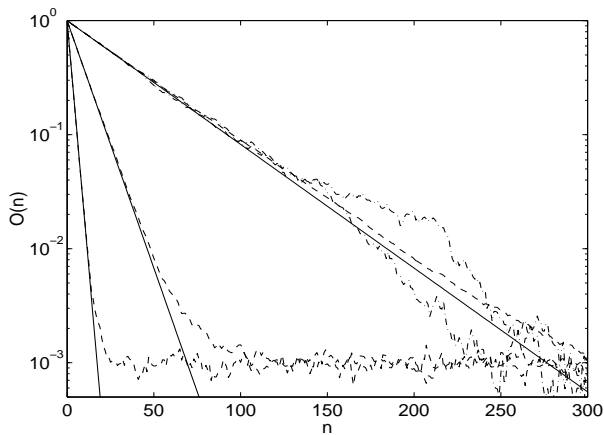, height = 5.8cm, width = 8cm}
\caption{\label{fig:cue} Fidelity decay for $U_{\text{CUE}}$ 
averaged over 50 computational basis states (dash lines) 
is in excellent agreement with the golden rule decay 
(\ref{eqn:fgr}) 
and the RMT rate (\ref{eqn:Gammacbr}) 
for $\delta = (0.1,0.2,0.4)$ (solid lines top to bottom). 
Chain lines show the fidelity decay for two typical 
computational basis states in the case $\delta = 0.1$.}
\end{figure}

While a CUE map may be generated on a QIP using the 
gate decomposition devised in Ref.~\cite{Zyc}, 
for our numerical study we construct $U=U_{\text{CUE}}$ 
directly from the eigenvectors of a random Hermitian matrix. 
Since computational basis states are easiest to prepare 
in the QIP setting, we consider the fidelity decay 
for both single computational basis states and averages over 
50 such states.  The behaviour of the fidelity decay 
for a matrix typical of CUE is displayed in Fig.~\ref{fig:cue}.  
The three perturbation values displayed in the figure 
are chosen near the onset of the non-perturbative regime ($\delta > 0.1 $) 
and it is evident that the fidelity decay even for 
{\em individual} computational basis 
states exhibits FGR decay (\ref{eqn:fgr}) at the expected rate.
The FGR decay persists for a 
time-scale $\Gamma^{-1}\log(N)$ 
until saturation at a time-average that decreases 
as $1/N$. 

\begin{figure}
\epsfig{file=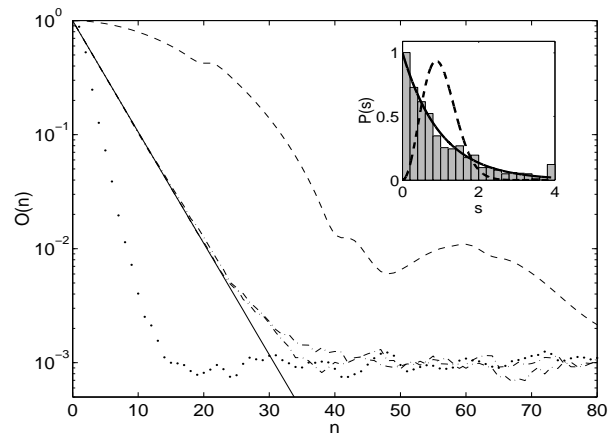, height = 5.8cm, width = 8cm}
\caption{ \label{fig:gue} Fidelity decay for $U_{\text{GUE}}$ averaged 
over 50 computational basis states 
with $\delta = 0.3$  for 
$\tau = 0.001$ (dashed line),  
$\tau = 0.01$ (dotted line) 
and $\tau = 0.1$ and $100$ (chain lines), compared to 
the FGR/RMT prediction (solid line). Inset: Spacing distribution 
for $U_{\text{GUE}}$ for $\tau =100$ compared to 
Poissonian (solid line) and 
Wigner-Dyson (dashed line) distributions.} 
\end{figure}

We next consider $O(n)$ for the Gaussian unitary ensemble 
(GUE) in order to clarify the relationship between the FGR decay 
and the distinct statistical features of RMT that represent 
signatures of quantum chaos. 
The GUE consists of Hermitian matrices with independent elements 
drawn randomly with respect to the unique unitarily-invariant 
measure \cite{Haake01}.  GUE forms the relevant RMT model 
for the important class of chaotic or complex 
autonomous Hamiltonian systems 
that are unrestricted by any additional symmetries.  
We may examine the sensitivity to perturbations for the GUE 
by constructing the  
unitary operator $U_{\text{GUE}} = \exp(-i H_{\text{GUE}} \tau )$, 
where $\tau$ is a time-delay between perturbations. We consider the same 
perturbation 
as for the CUE case.  For sufficiently small $\tau$ the 
propagator approaches identity and the overlap decay 
is dominated by the perturbation operator, 
$O(n) = |\langle \exp( - i \delta V) \rangle|^2 + O(\delta^2 \tau^2 n^2)$.  
This behaviour is demonstrated in Fig.~\ref{fig:gue} for $\tau = 0.001$ 
and $\tau=0.01$ and with $\delta = 0.3$.
For larger values of $\tau$ the fidelity decay under $U_{\text{GUE}}$ 
obeys the FGR with the RMT rate (\ref{eqn:Gammacbr}). 

The important point is that for sufficiently large 
$\tau$ the eigenphases of $U_{\text{GUE}}$ become 
spread pseudo-randomly in the interval $[0,2\pi)$. 
Under these conditions the 
eigenphases of the map $U_{\text{GUE}}$ exhibit the 
Poissonian spectral fluctuations that are characteristic 
of classically integrable (time-periodic) systems. 
We checked the nearest-neighbor spacing distribution 
of $U_{\text{GUE}}$ and 
found that for $\tau = 100$ the statistics are in excellent 
agreement with the Poissonian distribution $P(s) \propto \exp(-s)$ 
(see inset to Fig.~\ref{fig:gue}). 
However, the eigenvectors of $U_{\text{GUE}}$ 
are random (by construction) and 
independent of $\tau$  (for finite $\tau$). 
From these observations it is clear that the presence of 
Wigner-Dyson spectral 
fluctuations in the implemented $U$, which comprises  
the only basis-independent criterion of quantum chaos, 
is not actually necessary for the onset of exponential 
(FGR) decay at the rate (\ref{eqn:Gamma}). 
This suggests that it is the RMT statistics 
of the eigenvectors of $U$ 
that lead to the FGR decay with the RMT rate.  

\begin{figure}
\epsfig{file=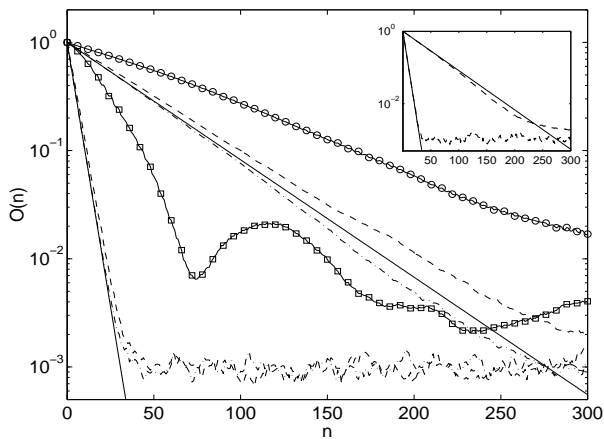, height = 5.8cm, width = 8cm}
\caption{ \label{fig:qkt} 
Decay of $O(n)$ for the kicked top in chaotic regime ($k=12$) 
averaged over 50 computational basis states, with   
the perturbation eigenbasis mapped to the eigenbases of 
$J_z$ (dash lines) and $J_y$ (chain lines) compared to 
the FGR decay (solid lines) at the RMT rate (\ref{eqn:Gammacbr}) for 
$\delta = (0.1,0.3)$ (top to bottom).  
Lines with circles and squares are for 
the regular kicked top ($k=1$) with the perturbation 
eigenbasis tied to the $J_z$ and $J_y$ 
coordinate bases respectively (for $\delta = 0.1$).  
Inset: Average fidelity decay for regular kicked top ($k=1$),   
when the qubit perturbation is 
in a random eigenbasis (\ref{eqn:rcbr})
and with $\delta = (0.1,0.3)$ (dashed lines),   
compared to the FGR/RMT rate (solid lines).}
\end{figure}

We next consider the fidelity decay for the quantum kicked 
top, which is an exemplary dynamical model of quantum 
chaos \cite{HKS87,Haake01}. 
The kicked top is a unitary map $U_{\text{QKT}} = \exp(-i \pi J_y /2) 
\exp(- i k J_z^2 /j)$ acting on the Hilbert space of dimension 
$N = 2j+1$ associated with an 
irreducible representation of the angular momentum operator $\vec{J}$. 
In previous fidelity decay and LDOS studies the choice of 
perturbation has usually been tied to a physical 
coordinate of the system $U$. 
We first follow this convention and identify 
the eigenbasis of the perturbation (\ref{eqn:cbr}) 
with the eigenbasis $| m_j \rangle$  of the system coordinate  
$J_z$ (where $m_j = \{j, \dots, -j\} $). 
In Fig.~\ref{fig:qkt} we compare the fidelity decay 
for the chaotic and regular regimes of the kicked top 
for averages over 50 initial computational basis states.  
The fidelity decay for the chaotic top ($k=12$) 
is well described by the FGR prediction (\ref{eqn:fgr}) 
and the RMT rate (\ref{eqn:Gammacbr}), 
whereas the regular top  ($k = 1$) 
shows a slower {\em non-exponential} decay rate. 
Similarly, if we associate the perturbation eigenbasis 
with the basis of the $J_y$ coordinate, 
the fidelity decay for the chaotic top remains 
in agreement with the RMT rate and the regular top 
again exhibits non-exponential decay, though in this case with a 
{\em faster} decay than the RMT rate (\ref{eqn:Gammacbr}). 
However, we now demonstrate that an exponential decay at the RMT rate 
arises even for the regular kicked top when the qubit 
perturbation (\ref{eqn:cbr}) is diagonal in 
a generic basis relative to the eigenbasis of $U_{\text{QKT}}$. 
Specifically, we leave the perturbation 
eigenvalue spectrum unchanged but set 
\begin{equation}
\label{eqn:rcbr}
U_p = T \left[ \Pi_j \exp(-i\delta \sigma_z^j /2 ) \right] T^{-1}, 
\end{equation}
where $T$ is drawn from CUE. As demonstrated in the inset 
to Fig.~\ref{fig:qkt}, under this type of perturbation the fidelity 
decay  for the regular top is indistinguishable from that of the  
chaotic top and is very accurately 
described by the FGR at the RMT rate (\ref{eqn:Gammacbr}). 

The sensitive dependence of the type of fidelity decay on the 
eigenbasis of the applied perturbation 
suggests a close connection with the basis-dependence of 
the eigenvector statistics of classically 
regular quantum models.  
Expressed in a generic quantum basis,  
the eigenvectors of any quantized classical system $U$ 
will have randomly (Gaussian) distributed components.  
In contrast, {\em in the eigenbases of the system coordinates} 
the components of classically chaotic and integrable systems 
are known to be different \cite{IZR87KMH88HZ90}, 
with the former Gaussian distributed 
and the latter exhibiting substantial deviation from 
the canonical Gaussian distribution. 
In light of this connection, in the case of quantized classical models 
we infer that exponential (non-exponential) fidelity decay 
can be correlated with the presence (absence) of characteristic RMT  
spectral fluctuations in the unperturbed system 
{\em provided that} the applied 
perturbation commutes with a system coordinate.  


In summary, we have shown that the fidelity decay 
may be measured efficiently on a QIP. 
We then examined 
which statistical properties of the unperturbed system  
determine the type and rate of the decay. 
In the case of random unitary and Hermitian matrices, 
as well as a classically chaotic dynamical model, we have 
shown that 
the fidelity decays exponentially with a characteristic rate 
given precisely by the variance of the perturbation's 
eigenspectrum. The occurrence of 
the exponential decay is not directly dependent on the 
Wigner-Dyson fluctuations of the unperturbed spectrum, but does  
depend sensitively on the RMT statistics of the system 
eigenvectors in the eigenbasis of the applied perturbation. 
Hence, the fidelity decay for both classically regular 
and chaotic dynamical systems is given by the FGR under all but a 
small subset of unitary perturbation operators. 
In the case of classical models, we conclude that the fidelity decay 
provides a reliable indicator of RMT statistics (quantum chaos) in the 
unperturbed system only when the applied perturbation is restricted 
to the subset of perturbations that commute with a classical coordinate.

We are grateful to K. Zyczkowski for suggesting the relevance 
of the eigenvector statistics, Ph.~Jacquod, E.M.~Fortunato, 
C.~Ramanathan, and T.F.~Havel for 
helpful discussions, and DARPA and the NSF for financial support. 


\begin{thebibliography}{99}
\bibitem{Lloyd}
S.~Lloyd, Science {\bf 273}, 1073 (1996).  
\bibitem{ZalkaWeisner}
C.~Zalka, Proc.~Roy.~Soc.~London, Ser.~A {\bf 454}, 313 (1998); 
S.~Weisner, quant-ph/9603028 (1996). 
\bibitem{Shep}
B.~Georgeot and D.L.~Shepelyansky, Phys. Rev. Lett. {\bf 86}, 2890 (2001).
\bibitem{Schack}
R.~Schack, Phys.~Rev.~A {\bf 57}, 1634 (1998). 
\bibitem{Casati}
G.~Benenti, G.~Casati, S.~Montangero, and D.L.~Shepelyansky, 
Phys.~Rev.~Lett.~{\bf 87}, 227901 (2001).
\bibitem{Yaakov}
Y.S.~Weinstein, S.~Lloyd, 
J.~Emerson, and D.~Cory, Phys. Rev. Lett. in press (2002).
\bibitem{BT77BGS84}
M.V.~Berry and M.~Tabor, Proc. Roy. Soc. Lond. {\bf A356}, 375 (1977);
O. Bohigas, M.J. Giannoni, C. Schmit, Phys. Rev Lett. {\bf 52}, 1 (1984). 
\bibitem{IZR87KMH88HZ90}
F.M.~Izrailev, Phys.~Lett.~{\bf 125A}, 250 (1987); 
M. Kus, J. Mostowski, and F. Haake, J. Phys. A: 
Math. Gen. {\bf 21}, L1073 (1988); 
F. Haake and K. Zyczkowski, Phys. Rev. A {\bf A42}, 1013 (1990). 
\bibitem{Peres}
A. Peres, Phys. Rev. A {\bf 30}, 1610 (1984). 
\bibitem{JP01Cucchietti01}
R.A. Jalabert and H.M. Pastawski, Phys. Rev. Lett. {\bf 86}, 2490 (2001); 
F. Cucchietti, C.H. Lewenkopf, E.R. Mucciolo, H. Pastawski
and R.O. Vallejos, nlin.CD/0112015.
\bibitem{Jacquod01}
Ph. Jacquod, P.G. Silvestrov, C.W.J. Beenakker, Phys. Rev. E {\bf 64}, 
055203 (2001).
\bibitem{Casati01}
G. Benenti and G. Casati, quant-ph/0112060. 
\bibitem{Prosen02}
T. Prosen and M. Znidaric, J. Phys. A {\bf 35}, 1455 (2002).
\bibitem{Wigner}
E.P. Wigner, Ann. Math. {\bf 62}, 548 (1955); {\bf 65}, 203 (1957).
\bibitem{FCIC96}
Y.V. Fyodorov, O.A. Chubykalo, F.M. Izrailev, and G. Casati, Phys.~Rev.~
Lett.~{\bf 76}, 1603 (1996).
\bibitem{Wisniacki02}
D. Wisniacki, E. Vergini, H. Pastawski, and F. Cucchietti, nlin.CD/0111051.
\bibitem{BGI98}
F. Borgonovi, I Guarneri, and F.M. Izrailev, Phys. Rev. E {\bf 57}, 5291 (1998).
\bibitem{Cory99}
C.H. Tseng, S. Somaroo, Y. Sharf, E. Knill, R. Laflamme, T.F. Havel, and 
D. Cory, Phys. Rev. A {\bf 61}, 012302 (1999).
\bibitem{Fortunato}
L. Viola, E. Fortunato, M. Pravia, E. Knill, R. Laflamme, and D. Cory, 
Science {\bf 293}, 2059 (2001).  
\bibitem{Haake01}
F. Haake, {\it Quantum Signatures of Chaos} (Springer, New York, 2001).
\bibitem{Zyc}
K. Zyczkowski and M. Kus, J. Phys. A {\bf 27}, 4235 (1994). 
\bibitem{HKS87}
F.~Haake, M.~Kus, and R.~Scharf, Z.~Phys.~B.~{\bf 65}, 381 (1987).
\end{thebibliography}
\end{document}